\documentclass[twocolumn,amsmath,amssymb]{revtex4}

\usepackage{graphicx}
\usepackage{graphics}
\usepackage{dcolumn}
\usepackage{bm}

\def\BEq{\begin{equation}}
\def\EEq{\end{equation}}
\def\BEqA{\begin{eqnarray}}
\def\EEqA{\end{eqnarray}}
\def\BEn{\begin{enumerate}}
\def\EEn{\end{enumerate}}
\def\BWT{\begin{widetext}}
\def\EWT{\end{widetext}}

\def\a{\alpha}

\def\bra{\langle}

\def\ket{\rangle}

\begin{document}


\title{Ground state of an exciton in a three-dimensional parabolic quantum dot: 
convergent perturbative calculation}

\author{Andrei Galiautdinov}
\email{ag@physast.uga.edu}
 \affiliation{
Department of Physics and Astronomy, 
University of Georgia, Athens, Georgia 30602, USA}

\date{\today}

\begin{abstract}
Working in the effective-mass approximation, we apply a powerful convergent 
perturbative technique of Turbiner's to the calculation of the 
ground state energy and the wave function of an exciton confined to a 
three-dimensional parabolic quantum dot. Unlike the usual 
Rayleigh-Schr\"{o}dinger perturbation theory, Turbiner's approach works 
well even in the regime of strong coupling and does not require the knowledge 
of the full solution to the undisturbed problem. The second-order convergent 
calculation presented below is in excellent agreement with the results of exact 
numerical simulations for a wide range of system's confinement parameters.
\end{abstract}


\maketitle


\section{Introduction}

A quantum dot is a semiconductor nanocrystal consisting of $10^3$ to $10^5$ 
atoms and having a typical size of $2$ to $10$ nanometers \cite{Kastner1993}. 
The energy spectrum of a quantum dot is fundamentally different from that of 
a bulk semiconductor. A particle (such as an electron or a hole) in a nanocrystal 
behaves as if it was confined to a three-dimensional potential well, which makes 
particle's energy levels discrete and strongly dependent on the dot's size 
\cite{Ekimov1981, Ekimov1984, Ekimov1986}. 
That leads to various quantum size effects which, in turn, give the dot its 
unique optoelectric properties \cite{Kayanuma1988, Ciftja2013}.

Of particular interest to us is the influence of the confining potential 
on the ground state of an exciton \cite{Knox1963} --- an elementary excitation of 
the nanocrystal consisting of an electron and a hole interacting with each other 
by way of the usual electrostatic potential. Over the last thirty years, 
numerous approaches to the calculation of the ground state of the exciton 
have been employed by different authors. Chief among them are: 
the power series approach (also known as Frobenius method) 
\cite{Zhu1990, Hassanabadi2009_1, Hassanabadi2009_2}, 
the standard Rayleigh-Schr\"{o}dinger perturbation theory 
\cite{Merkt1990, Babich1992, Ciftja2004}, the $1/N$ expansion 
\cite{El-Said1994, McKinney2000},
the expansion in electron-hole product states method \cite{Garm1996},
the variational method \cite{Adamowski2000, Ciftja2004, 
JahanK2015, Semina2016},
the spin-density-functional-theory approach \cite{Hirose2002},
the analytical iteration method with trial wave function \cite{Ikhdair2011}, 
various numerical matrix diagonalization techniques 
\cite{Hu1990, Kaputkina1998, Ciftja2004, Semina2016}, and the use of the 
finite-element numerical solvers \cite{Moskalenko2007}.
Here we propose yet another approach --- a powerful convergent 
perturbative method due to Turbiner 
\cite{Turbiner1980, Turbiner1981, Turbiner1984},
which works well even in the regime of strong coupling and does 
not require the knowledge of the full solution to the undisturbed 
problem.

In the effective-mass model, the exciton in a quantum dot is described by the 
Hamiltonian \cite{Que1992},
\begin{align}
\label{eq:H_exciton_1}
H_{\rm exciton}&=-\frac{\hbar^2}{2m^{*}_e}\nabla^2_{e}
-\frac{\hbar^2}{2m^{*}_h}\nabla^2_{h}
-\frac{e^2}{\epsilon |{\bf r}_{e}-{\bf r}_{h}|}
\nonumber \\
&\quad
+\frac{1}{2}m^{*}_{e}\omega^2 r_e^2
+\frac{1}{2}m^{*}_{h}\omega^2 r_h^2,
\end{align}
where $m^{*}_{e,h}$ and ${\bf r}_{e,h}$ are 
the (effective) masses and positions of the
electron and the hole, $r_{e,h} \equiv |{\bf r}_{e, h}|$, and
$\epsilon$ is the dielectric constant of the ambient material. 
The last two terms in (\ref{eq:H_exciton_1}) represent the leading 
(harmonic) part of a more realistic confining potential. It is assumed 
that the frequency, $\omega$, of small oscillations near the bottom of 
the well is independent of the particle's mass, which is similar to the 
case of a particle oscillating at the bottom of a frictionless bowl under 
the action of uniform gravity, familiar from introductory mechanics. 
The name ``parabolic'' used to describe the quantum dot comes from 
the shape of this confining potential \cite{Que1992, Kaputkina1998}. 

Introducing the center-of-mass and the relative coordinates,
\BEq
{\bf R}=\frac{m^{*}_e {\bf r}_{e} 
+ m^{*}_h {\bf r}_{h}}{m^{*}_e+m^{*}_h},
\quad
{\bf r}={\bf r}_{e} -{\bf r}_{h},
\EEq
we re-write system's Hamiltonian as the sum,
\BEq
H_{\rm exciton}= H_{\rm COM} + H_{\rm relative},
\EEq
where
\begin{align}
H_{\rm COM}&=-\frac{\hbar^2}{2M}\nabla^2_{{\bf R}}
+\frac{M\omega^2 R^2}{2},
\\
H_{\rm relative}&=
-\frac{\hbar^2}{2m}\nabla^2_{{\bf r}}
-\frac{e^2}{\epsilon r}
+\frac{m\omega^2 r^2}{2},
\end{align}
and
\BEq
M=m^{*}_e + m^{*}_h,
\quad
m= \frac{m^{*}_e  m^{*}_h}{{m^{*}_e+m^{*}_h}}.
\EEq
The system's dynamics, thus, separates into the center-of-mass 
motion of the exciton as a whole, which is described by $H_{\rm COM}$ 
whose solution is trivial, and the motion of a fictitious particle of mass 
$m$ described by $H_{\rm relative}$ with potential 
\BEq
V=-\frac{e^2}{\epsilon r}+\frac{m\omega^2 r^2}{2}.
\EEq
Upon setting
\BEq
\a \equiv me^2/(\epsilon \hbar^2),
\quad
b \equiv m^2 \omega^2/\hbar^2,
\EEq
the potential assumes the general form,
\begin{align}
\label{eq:V}
V=-\frac{2\alpha}{r} + b r^2,
\end{align}
with the corresponding Schr\"{o}dinger's equation being 
\begin{align}
\label{eq:SchEq01}
\nabla^2 \psi + (E-V)\psi = 0,
\end{align}
where we dropped the index $_{\bf r}$ on the nabla operator 
and re-scaled the energy by the factor of $2m/\hbar^2$ for future 
convenience. Our goal will be to find the ground state of the relative 
Hamiltonian, $H_{\rm relative}$, for {\it any} values of $\a$ and $b$.
[For the possibility of having analytical solutions for some
particular, denumerably infinite set of oscillator frequencies ---
the so-called quasi-exactly-solvable problem --- see 
Refs.\ \cite{Taut1993, Turbiner1994}.]

\section{Formal perturbative calculation}
\label{sec:formalTheory}

To determine the ground state of the exciton, we use the convergent 
perturbation theory developed by Turbiner 
\cite{Turbiner1980, Turbiner1981, Turbiner1984}. 
Turbiner's theory consists of two parts. 
The first part (used in this section) deals with the formal 
development of the perturbative scheme applicable to any potential, 
$V$, that can be split into a formal sum,
\BEq
V=V_0+\lambda V_1, 
\EEq
of its zeroth-order (unperturbed) component, $V_0$, and the 
perturbation, $\lambda V_1$, where $\lambda$ is the perturbation 
parameter. At this stage, no specific requirements are imposed on 
the perturbation other than that it has to be small in some  reasonable 
sense. For example, in the case of a quantum dot, we may split the 
potential in (\ref{eq:V}) according to
\BEq
\label{eq:weak-confinement}
V_0=-\frac{2\alpha}{r}, \quad \lambda V_1 =  b r^2,
\EEq
which would correspond to the so-called
weak-confinement regime \cite{Que1992}, with $b$ playing the role of 
the perturbation parameter. Alternatively, we could choose
\BEq
\label{eq:strong-confinement}
V_0=b r^2, \quad \lambda V_1 =  -\frac{2\alpha}{r},
\EEq
and work in the strong-confinement regime \cite{Que1992} with small 
parameter $\alpha$. Regardless of the $V$-split, the ground state wave 
function, $\psi$ (here assumed to be nodeless; for the possibility of having 
a nodal ground state see \cite{Lee2014}), 
is sought in the form
\BEq
\psi ({\bf r}) = e^{-\phi({\bf r})},
\EEq
with $\phi$ being a power series in $\lambda$,
\BEq
\phi = \phi_0 + \lambda \phi_1 + \lambda^2 \phi_2 + \dots,
\EEq
and, similarly, for the ground state energy,
\BEq
E = E_0 + \lambda E_1 + \lambda^2 E_2 + \dots.
\EEq 
The actual perturbative calculation is not applied to $\phi$ directly, 
however. Instead, Turbiner
introduces an auxiliary vector field, ${\bf y},$ defined by
\BEq
{\bf y} = -\nabla \psi/\psi = \nabla \phi,
\EEq
which satisfies the non-linear differential equation (hence the name 
``nonlinearization procedure'' used in his papers),
\BEq
\nabla \cdot {\bf y} - {\bf y}^2 = E-V,
\EEq
with the boundary condition,
\BEq
|\psi^2{\bf y}|\rightarrow 0, \quad r\rightarrow \infty,
\EEq
which corresponds to the requirement that the probability current at 
infinity must vanish.
The series expansion for ${\bf y}$,
\BEq
{\bf y} = {\bf y}_0 + \lambda {\bf y}_1 + \lambda^2 {\bf y}_2 + \dots,
\EEq
is then found using the iteration scheme,
\begin{align}
\label{eq:iterationScheme01}
\nabla\cdot {\bf y}_{n}  - 2{\bf y}_0\cdot{\bf y}_{n} &=E_{n}-Q_{n}, 
\quad n = 1,2,3, \dots,
\end{align}
where
\begin{align}
\label{eq:22}
{\bf y}_0 &= -\frac{\nabla \psi_0}{\psi_0},
\\
\label{eq:23}
Q_1 &= V_1, 
\\
\label{eq:24}
E_{n} &= \frac{\int \psi_0^2 \, Q_{n} \, dV}{\int \psi_0^2 \, dV}, 
\quad n = 1,2,3, \dots,
\\
\label{eq:25}
Q_n &=-\sum_{i=1}^{n-1}{\bf y}_i\cdot{\bf y}_{n-i}, 
\quad n = 2,3, \dots,
\end{align}
with $dV$ being the volume element in the configuration space of the 
system. The remarkable feature of Turbiner's method, which makes it 
very different from the usual perturbation theory, is that it does not 
require the complete solution (all wave functions and the full spectrum) 
of the undisturbed problem.

In the spherically-symmetric case, all ${\bf y}_n$ have the form,
\BEq
\label{eq:26}
{\bf y}_n=y_n(r)\hat{\bf r}, \quad n=0,1,2,3,\dots,
\EEq
with $\hat{\bf r}={\bf r}/r$, and (\ref{eq:iterationScheme01}) reduces 
to a much simpler sequence of first order linear ordinary differential equations,
\BEq
\label{eq:27}
ry_n'=2(y_0r-1)y_n+r(E_n-Q_n), \quad n=1,2,3,\dots,
\EEq
each of which has the general solution,
\BEq
\label{eq:28}
y_n(r) = C_ne^{F(r)}+e^{F(r)}\int e^{-F(r)} [E_n-Q_n(r)]\,dr,
\EEq
with
\BEq
\label{eq:29}
F(r)=\int \frac{2(y_0r-1)}{r}\,dr,
\EEq
where the constants $C_n$ must be determined from the corresponding 
boundary conditions,
\BEq
\label{eq:30}
|\psi_0^2{y}_n|\rightarrow 0, \quad r\rightarrow \infty.
\EEq
Applying this scheme to the case of a weakly confined dot, 
Eq.\ (\ref{eq:weak-confinement}), we first choose
\BEq
\psi_0=e^{-\a r}, \quad E_0 = -\a^2,
\EEq
and then find, upon direct calculation,
\begin{align}
\label{eq:asymptotic01}
E &= -\a^2+\frac{3\,b}{\a^2}-\frac{129 \, b^2}{8\, \a^6}
+\frac{5451 \, b^3}{16 \, \a^{10}}
-\frac{6609975 \, b^4}{512 \, \a^{14}}+\dots,
\\
\label{eq:asymptotic02}
\phi
&=
\a r \, 
\bigg[
1+ \frac{b r }{6 \, \a^3}(3+\a r)
\nonumber \\
&\quad
-\frac{b^2 r }{480 \, \a^7}(1290+430\,\a r + 105 \,\a^2 r^2 
+ 12\,\a^3 r^3)
\nonumber \\
&\quad
+\frac{b^3 r }{6720 \, \a^{11}}(381570+127190\,\a r 
+ 34545 \,\a^2 r^2 
\nonumber \\
&\quad
+ 6804\,\a^3 r^3+910\,\a^4 r^4+60\,\a^5 r^5)+\dots
\bigg].
\end{align}
Similarly, in the strongly confined case, Eq.\ (\ref{eq:strong-confinement}), 
we set
\BEq
\psi_0=e^{-(\sqrt{b}/2) r^2}, \quad E_0 = 3\sqrt{b},
\EEq
and get
\begin{align}
E &= 3\sqrt{b}-\frac{4 {\a} \sqrt[4]{b}}{\sqrt{\pi }}
-\frac{4 {\a}^2}{\sqrt{\pi }\sqrt[4]{b} }
\int_0^{\infty}
e^{-\sqrt{b} r^2} 
 \nonumber \\
& \quad 
\times
\bigg\{
\frac{1}{r} \left[e^{\sqrt{b} r^2}\text{erfc}(\sqrt[4]{b} r)-1\right]
+\frac{2 \sqrt[4]{b}}{\sqrt{\pi }}
\bigg\}^2 dr
+\dots,
\\
\phi
&=
\frac{\sqrt{b}}{2} r^2
 \nonumber \\
& \quad 
+\frac{\a}{\sqrt{b}}
\int_0^r 
\bigg\{
\frac{1}{{\tilde r}^2}
\left[e^{\sqrt{b} {\tilde r}^2} \text{erfc}(\sqrt[4]{b} {\tilde r})-1\right]
+\frac{2 \sqrt[4]{b}}{\sqrt{\pi }} \frac{1}{{\tilde r}}
\bigg\}d{\tilde r}+\dots,
\end{align}
where
\begin{align}
\text{erf}(x)&=\frac{2}{\sqrt{\pi}}\int_0^x e^{-t^2}dt \nonumber \\
&\sim
\frac{2e^{-x^2}}{\sqrt{\pi}}\bigg(x+\frac{2x^3}{1·3}
+\frac{4x^5}{1·3·5}+\dots \bigg), \quad x \rightarrow 0,
\\
\text{erfc}(x)&= 1-\text{erf}(x)
\nonumber \\
&\sim 
\frac{e^{-x^2}}{\sqrt{\pi}}
\bigg(\frac{1}{x}-\frac{1}{2x^3}+\frac{3}{4x^5}-\dots\bigg), 
\quad x \rightarrow \infty,
\end{align}
are the error and complementary error functions.

The main problem with the formal perturbative scheme 
(be it the usual one, or the one described above) is evident from 
Eqs.\ (\ref{eq:asymptotic01}), (\ref{eq:asymptotic02}) describing 
the weakly confined quantum dot. The resulting series for the ground 
state energy and for the exponent of the wave function are not actually
convergent, but asymptotic. As with any asymptotic expansion, one has 
to exercise care when interpreting the calculated results. The asymptotic 
expansion, by its very nature, works best in the limit when perturbation 
approaches zero, in which case one may keep many terms in the series. 
However, when the perturbation increases, the expansion begins to break 
down and only a small number of terms may be kept, thus reducing the 
accuracy of the final result. One describes this situation formally by 
saying that the series has zero radius of convergence. This is very 
different from a series whose radius of convergence is different from 
zero: in that case, the more terms one keeps, the better the resulting 
approximation. 

\section{Convergent approach}

The reason for the appearance of diverging series (\ref{eq:asymptotic01}) 
and (\ref{eq:asymptotic02}) in our calculation is that, for sufficiently 
large $r$, the perturbation potential, $V_1 =  b r^2$, 
becomes larger than the unperturbed potential, $V_0=-{2\alpha/r}$. 
Because of that, $V_1$ cannot be regarded as a ``good'' perturbation 
acting on the unperturbed system.

To develop a perturbative scheme that is {\it convergent} (and works 
even for strong perturbations), Turbiner formulates the following 
prescription: choose $V_0$ in such a way that it reproduces as many 
of the characteristic properties of the full potential $V$ as possible, 
such as, e.\ g., its singular points, asymptotic behavior, etc. 

\begin{figure}[!ht]
\includegraphics[angle=0,width=1.00\linewidth]{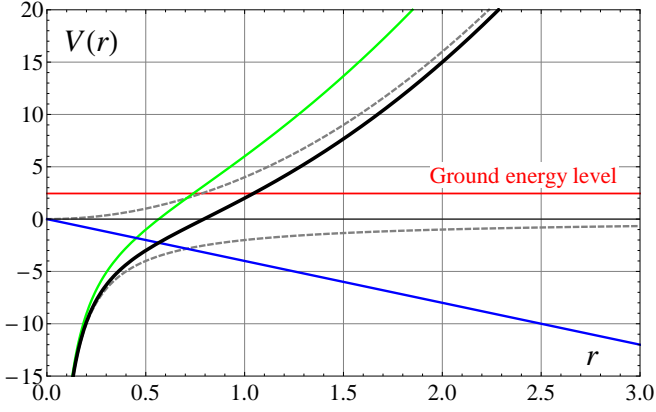}
\caption{ \label{fig:1} 
(Color online.) 
Black curve: Graph of excitonic potential energy, $V(r)$, Eq.\ (\ref{eq:V}), 
at $\a=1$ and $b=4$. Gray dashed curves: the Coulomb and harmonic 
oscillator potentials comprising $V$. Green curve: zeroth-order potential, 
$V_0$, as given in Eq.\ (\ref{eq:V0_convergent}). Blue curve: ``good'' 
perturbation, $V_1$, as given in Eq.\ (\ref{eq:V1_convergent}). 
Red curve: ground level, with energy $E_{\rm num}=2.4474$ 
(numerically calculated value).
Second-order convergent perturbative calculation gives $E=2.4632$; 
see below.
}
\end{figure}

To see how this prescription works in the case of a quantum dot 
subjected to an arbitrary parabolic potential (\ref{eq:V}) we first choose 
\BEq
\label{eq:psi0_convergent}
\psi_0 = e^{-\a r - \sqrt{b}r^2/2}
\EEq
as the zeroth-order approximation to the ground state wave function.
This is a very reasonable choice, for in both critical limits, $r\rightarrow 0$ 
and $r\rightarrow \infty$, the behavior of $\psi_0$ matches the behavior 
of the ground state wave function of the corresponding dominating 
potential, $-{2\alpha/r}$ or $b r^2$, respectively. 
[Notice that (\ref{eq:psi0_convergent}) is similar in form to the 
``variational wavefunction'' proposed in Ref.\ \cite{JahanK2015}, 
though in that case $\a$ and $b$ were treated as the variational 
parameters.]
The corresponding zeroth order potential, $V_0$, and the ground state 
energy, $E_0$, are then readily found by calculating the quantity
\begin{align}
\label{eq:quantity_convergent}
\frac{\nabla^2 \psi_0}{\psi_0}
&= \left(-\frac{2\a}{r}+br^2+2\a\sqrt{b}r\right)
-\left(-\a^2+3\sqrt{b}\right),
\end{align}
which, on the basis of the zeroth order Schr\"{o}dinger equation, 
should be identified with the difference $V_0-E_0$. 
From (\ref{eq:quantity_convergent}), we immediately get
\begin{align}  
\label{eq:V0_convergent}
V_0&= -\frac{2\a}{r}+br^2+2\a\sqrt{b}r,
\\
\label{eq:E0converging}
E_0&=-\a^2+3\sqrt{b},
\end{align}  
as well as the ``good'' perturbation (see Fig.\ \ref{fig:1}),
\BEq 
\label{eq:V1_convergent}
\lambda V_1 = -2\a\sqrt{b}r.
\EEq
The asymptotic behavior of the found $V_0$ correctly reproduces 
the behavior of the full potential $V$ at its critical points, as required 
by Turbiner's prescription. Additionally, the perturbation $V_1$ never
dominates the zeroth-order $V_0$ for any value of $r$.

\begin{figure}[!ht]
\includegraphics[angle=0,width=1.00\linewidth]{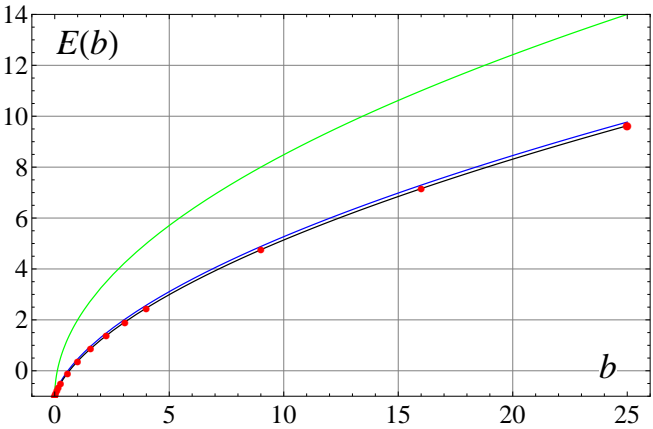}
\caption{ \label{fig:2} 
(Color online.) 
Ground state energy, $E$, of the exciton, Eq.\ (\ref{eq:V}), at $\a=1$ 
as a function of the parabolicity parameter, $b$. Green curve: plot of 
zeroth-order approximation, $E\approx E_0$, using Eq.\ (\ref{eq:E0converging}). 
Blue curve: plot of the first-order approximation, $E\approx E_0+E_1$, using 
Eqs.\ (\ref{eq:E0converging}) and (\ref{eq:E1converging}).
Black curve: plot of the second-order approximation, 
$E\approx E_0+E_1+E_2$, using Eqs.\ (\ref{eq:E0converging}), 
(\ref{eq:E1converging}), and (\ref{eq:E2converging}). 
Red dots: exact numerical simulation.   
For example, at $b=4$, we have $E_0=5$, $E_1=-2.42345$, 
$E_2=-0.113349$, $E=2.4632$, and $E_{\rm num}=2.4474$.
}
\end{figure}

\begin{figure}[!ht]
\includegraphics[angle=0,width=1.00\linewidth]{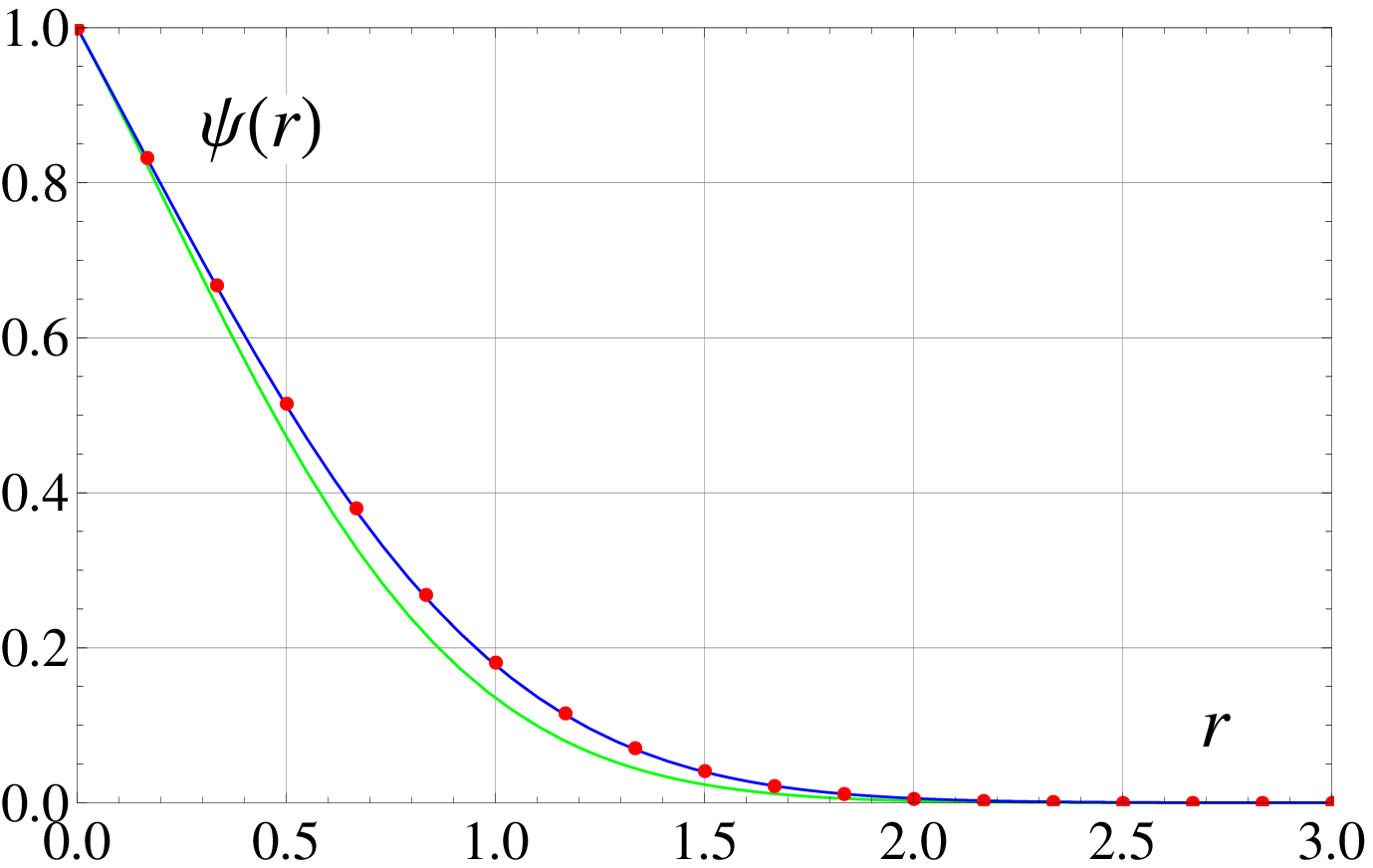}
\caption{ \label{fig:3} 
(Color online.) 
Ground state wave function, $\psi(r)$, of the exciton, 
Eq.\ (\ref{eq:V}), at $\a=1$ and $b=4$. 
Green curve: plot of the zeroth-order approximation, 
$\psi_0$, using Eq.\ (\ref{eq:psi0_convergent}). 
Blue curve: plot of the first-order approximation, 
$\psi = e^{-\phi(r)}$, with $\phi$ given in 
Eq.\ (\ref{eq:psi01Convergent}).
Red dots: exact numerical simulation.   
}
\end{figure}

From here on, the calculation proceeds in accordance with the general 
scheme outlined in Section \ref{sec:formalTheory}.
Using (\ref{eq:22}), (\ref{eq:23}), (\ref{eq:24}), (\ref{eq:25}),  
and (\ref{eq:28}), we find
\BWT
\begin{align}
y_1
&=
\frac{1}{r^2}
\left[
C_1 e^{r \left(2 \a+\sqrt{b} r\right)}
-\frac{
\a \left(
\sqrt{\pi } e^{\frac{\a^2}{\sqrt{b}}} 
\text{erfc}\left(\frac{\a}{\sqrt[4]{b}}\right) 
\left(2 \a^2 r^2+2 \a r+\sqrt{b} r^2+1\right)
+\sqrt{\pi } e^{\frac{\left(\a+\sqrt{b} r\right)^2}{\sqrt{b}}} 
\text{erf}\left(\frac{\a+\sqrt{b} r}{\sqrt[4]{b}}\right)
-2 \sqrt[4]{b} r (\a r+1)
\right)
}{
\sqrt{\pi } \left(2 \a^2+\sqrt{b}\right) e^{\frac{\a^2}{\sqrt{b}}} 
\text{erfc}\left(\frac{\a}{\sqrt[4]{b}}\right)-2 \a \sqrt[4]{b}
}
\right],
\end{align}
and, upon setting
\BEq
C_1=\frac{\sqrt{\pi } {\a} e^{\frac{a^2}{\sqrt{b}}}}
{\sqrt{\pi } \left(2 \a^2+\sqrt{b}\right) 
e^{\frac{\a^2}{\sqrt{b}}} 
\text{erfc}\left(\frac{\a}{\sqrt[4]{b}}\right)
-2 {\a} \sqrt[4]{b}},
\EEq
get
\begin{align}
y_1&=
-\frac{
{\a} \left[
\sqrt{\pi } e^{\frac{\a^2}{\sqrt{b}}} 
\text{erfc}\left(\frac{\a}{\sqrt[4]{b}}\right) 
\left(2 \a^2 r^2+2 {\a} r+\sqrt{b} r^2+1\right)
-\sqrt{\pi } e^{\frac{\left(\a+\sqrt{b} r\right)^2}{\sqrt{b}}} 
\text{erfc}\left(\frac{\a+\sqrt{b} r}{\sqrt[4]{b}}\right)
-2 \sqrt[4]{b} r({\a} r+1)
\right]
}{r^2 
\left[
\sqrt{\pi } \left(2 \a^2+\sqrt{b}\right) e^{\frac{\a^2}{\sqrt{b}}} 
\text{erfc}\left(\frac{\a}{\sqrt[4]{b}}\right)-2 {\a} \sqrt[4]{b}
\right]
},
\end{align}
\EWT
which vanishes at the origin and satisfies the boundary condition, (\ref{eq:30}), 
at infinity. We then get the first-order correction to the ground state energy,
\BWT
\begin{align}  
\label{eq:E1converging}
E_1 
&= 
\frac{
2 \a 
\bigg[
\sqrt{\pi }  
\a
\left(2 \a^2+3 \sqrt{b}\right) 
e^{\frac{\a^2}{\sqrt{b}}} \text{erfc}\left(\frac{\a}{\sqrt[4]{b}}\right)
-2 \sqrt[4]{b} \left(\a^2+\sqrt{b}\right)
\bigg]
}{
\sqrt{\pi } 
\left(2 \a^2+\sqrt{b}\right) 
e^{\frac{\a^2}{\sqrt{b}}} \text{erfc}\left(\frac{\a}{\sqrt[4]{b}}\right)
-2 {\a} \sqrt[4]{b}
},
\end{align} 
the second-order correction,
\begin{align} 
\label{eq:E2converging}
E_2 &= 
\frac{-4 \a^2 b^{5/4}}
{
\bigg[
\sqrt{\pi }  
\left(2 \a^2+\sqrt{b}\right)
e^{\frac{\a^2}{\sqrt{b}}} \text{erfc}\left(\frac{\a}{\sqrt[4]{b}}\right)
-2 {\a} \sqrt[4]{b}
\bigg]^3
} 
\int_0^{\infty}
\frac{
 e^{-r \left(2 \a+\sqrt{b} r\right)} 
}
{
r^2 
}
 \nonumber \\
& \quad 
\times
\bigg[
\sqrt{\pi } 
\left(2 \a^2 r^2+2 {\a} r+\sqrt{b} r^2+1\right)
e^{\frac{\a^2}{\sqrt{b}}} \text{erfc}\left(\frac{\a}{\sqrt[4]{b}}\right) 
-\sqrt{\pi } e^{\frac{\left(\a+\sqrt{b} r\right)^2}{\sqrt{b}}} 
\text{erfc}\left(\frac{\a+\sqrt{b} r}{\sqrt[4]{b}}\right)
-2 \sqrt[4]{b} r ({\a} r+1)
\bigg]^2
\, dr,
\end{align} 
and the exponent of the wave function in first order,
\begin{align} 
\label{eq:psi01Convergent}
\phi &=
{\a}r+\frac{\sqrt{b}}{2} r^2 
+
\frac{-\a}
{
\sqrt{\pi } \left(2 \a^2+\sqrt{b}\right) e^{\frac{\a^2}{\sqrt{b}}} 
\text{erfc}\left(\frac{\a}{\sqrt[4]{b}}\right)-2 \a \sqrt[4]{b}
}
\nonumber \\
&\quad
\times
\int_0^r
\frac{
\sqrt{\pi } 
e^{\frac{\a^2}{\sqrt{b}}} \text{erfc}\left(\frac{\a}{\sqrt[4]{b}}\right) 
\left(2 \a^2 {\tilde r}^2+2 {\a} {\tilde r}+\sqrt{b} {\tilde r}^2+1\right)
-\sqrt{\pi } 
e^{\frac{\left(\a+\sqrt{b} {\tilde r}\right)^2}{\sqrt{b}}} 
\text{erfc}\left(\frac{\a+\sqrt{b} {\tilde r}}{\sqrt[4]{b}}\right)
-2 \sqrt[4]{b} 
{\tilde r} ({\a} {\tilde r}+1)
}{
{\tilde r}^2 
}
\, d{\tilde r},
\end{align}
\EWT
with the corresponding results depicted in Figs.\ \ref{fig:2} and \ref{fig:3}. 
As seen from those figures, Turbiner's method works exceptionally well 
for a wide range of confinement parameters.

Notice that the found $E_1$ coincides with the first-order correction 
to the ground state energy calculated on the basis of the usual perturbation 
theory,
\BEq
E_1 = \frac{\bra \psi_0 | V_1 | \psi_0 \ket}{\bra \psi_0 | \psi_0 \ket}
\equiv \frac{\int_{0}^{\infty} V_1  \psi_0^2 \,  r^2 dr }
{\int_{0}^{\infty} \psi_0^2 \,  r^2 dr}.
\EEq
Notice also that the second-order correction, 
\BEq
E_2=-\frac{\int_{0}^{\infty} y_1^2  \psi_0^2 \, r^2 dr}
{\int_{0}^{\infty} \psi_0^2 \, r^2 dr},
\EEq
is always negative, as should have been expected. These two general
properties of Turbiner's scheme had already been pointed out in the original 
papers on the subject \cite{Turbiner1980, Turbiner1981, Turbiner1984}.

\section{Summary}
Assuming the effective mass model, we performed a 
convergent perturbative calculation of the ground state 
of an exciton confined by the parabolic potential of a 
three-dimensional quantum dot. No use of the full solution 
of the unperturbed problem has been made, except for the 
easily determined expressions for the unperturbed ground 
state energy and the wave function. For a wide range of 
system's parameters, our approximate solution is in very good 
agreement with the results of exact numerical simulations. 


\end{document}